# ZnO nanowires grown on $Al_2O_3$-$ZnAl_2O_4$ nanostructure using solid-vapor mechanism

Wiktoria Zajkowska[*], Jakub Turczyński[*], Bogusława Kurowska[*], Henryk Teisseyre[*,**], Krzysztof Fronc[*], Jerzy Dąbrowski[*] and Sławomir Kret[*]

[*] Institute of Physics Polish Academy of Sciences, Warsaw, Poland

[**] Institute of High Pressure Physics, Polish Academy of Sciences, Warsaw, Poland

Abstract

We present $Al_2O_3$-$ZnAl_2O_4$-ZnO nanostructure, which could be a prominent candidate for optoelectronics, mechanical and sensing applications. While ZnO and $ZnAl_2O_4$ composites are mostly synthesized by sol-gel technique, we propose a solid-vapor growth mechanism. To produce $Al_2O_3$-$ZnAl_2O_4$-ZnO nanostructure, we conduct ZnO:C powder heating resulting in ZnO nanowires (NWs) growth on sapphire substrate and $ZnAl_2O_4$ spinel layer at the interface. The nanostructure was examined with Scanning Electron Microscopy (SEM) method. Focused Ion Beam (FIB) technique enabled us to prepare a lamella for Transmission Electron Microscopy (TEM) imaging. TEM examination revealed high crystallographic quality of both spinel and NW structure. Epitaxial relationships of $Al_2O_3$-$ZnAl_2O_4$ and $ZnAl_2O_4$-ZnO are given.

1. Introduction

ZnO NWs find wide interest because of their remarkable physical properties which make them prominent candidates for optoelectronic[1], [2], nanomechanical devices[3] and sensors[4]. ZnO exhibits direct wide bandgap equal to 3.36 meV at room temperature (RT), its electron binding energy equals 60 meV (GaN – 25 meV, ZnSe – 26 meV), it has high radiation resistances and piezoelectric properties[5], [6]. On the other hand, the $ZnAl_2O_4$ spinel is a possible future air pollution remover, as it can degrade very toxic toluene[7]. In addition, it is reported to be a potential sensor due to the variation of luminescence with thermal history[8]. Due to the wide optical bandgap ($E_g$ = 3.8 eV) $ZnAl_2O_4$ can also be used in optoelectronics. Moreover, E. Chikoidze et. al. reported that different spinel from the same family: $ZnGa_2O_4$ is *p*-type widest bandgap ternary oxide (Eg = 5 eV)[9]. Such a prominent combination attracted scientists attention and resulted in $ZnO/ZnAl_2O_4$ nanostructures research. $ZnO/ZnAl_2O_4$ composites are studied as promising candidates for highly selective hydrogen gas sensing[10], long-term stable photocatalyst[11] or rapid dye degradation environmental applications[12].

ZnO and $ZnAl_2O_4$ composites are mostly synthesized by sol-gel[12] method, high doping of ZnO with $Al_2O_3$ or Al and longtime annealing[13], atomic layer deposition (ALD) technique[14] and solid-vapor or solid-solid ZnO-$Al_2O_3$ reactions. Jáger et. al.[15] reported lateral growth of $ZnAl_2O_4$ phase islands, then planar growth as a continuous layer at 700 °C using crystalline ZnO-amorphous $Al_2O_3$ bilayers. In general, the common way to obtain a spinel oxide, e. g. $ZnAl_2O_4$, are solid-state reactions: $AO + B_2O_3 \rightarrow AB_2O_4$ type. C. Gorla et. al.[16] studied the $ZnAl_2O_4$ formation process by solid reactions between (11-20) ZnO and (01-12) $Al_2O_3$. H. Fan et. al. indicated that $ZnAl_2O_4$ fabrication by reaction of solid ZnO with $Al_2O_3$ vapor is unique among other spinels ($Zn_2SiO_4$[17], $Zn_2GaO_4$[18]), because the growth process consists of diffusion of both oxygen and zinc atoms. It results in unilateral transfer of ZnO into the $ZnAl_2O_4$ layer[19]. We propose a method of $Al_2O_3$-$ZnAl_2O_4$-ZnO NWs crystalline structure fabrication using solid-vapor growth mechanism.

2. Experimental

Zinc oxide (Sigma-Aldrich ReagentPlus, powder grain size < 5 µm, 99,9%) and graphite (Supelco, powder grain size < 50 µm, 99,5%) were homogenized in a mortar in a 1: 1 molar ratio. After which, the powder was placed in a corundum crucible, covered with a sapphire substrate [11-20] and the corundum 75%-covering lid was placed on top of it. The amount of powder was adjusted to fill ca. 90% of the space under the substrate, then the powder was weighed. For the repeatability of the processes, the crucible filling degree was related to the powder mass. The system was placed in an open quartz tube in a heated tubular furnace. The temperature of the process was equal to 975 °C. After 15 minutes of heating, the furnace was turned off. The system cooled down to 700 °C, then it was removed from the furnace and quickly reached RT.

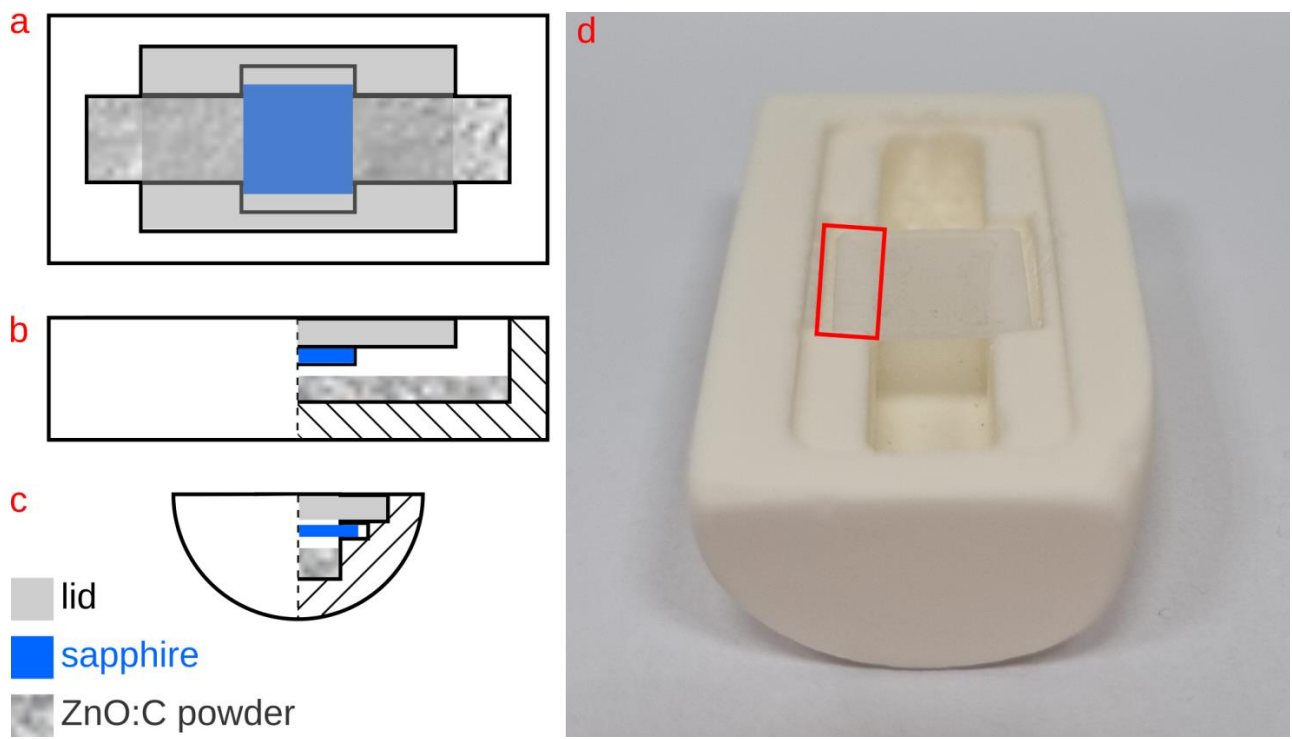

Figure 1. The set-up scheme: top (a), side (b) and front view (c). Sapphire substrate is placed on the crucible, the red rectangle indicates the substrate-crucible contact region (d).

3. Results and discussion

ZnO NWs morphology was examined using a FEI Helios Nanolab 600 SEM. In Figure 2 SEM images of the as-grown structures are presented. The ZnO NWs are densely distributed on the substrate, which is perpendicular to scanning electron beam (Figure 2a) and tilted by 45° (Figure 2b). Horizontal lines visible in Figure 2 result from beam shifts caused by charging of the semiconductor specimen. The ZnO NWs length is 50-70 µm and their diameters range from dozens of nm to 400 nm.

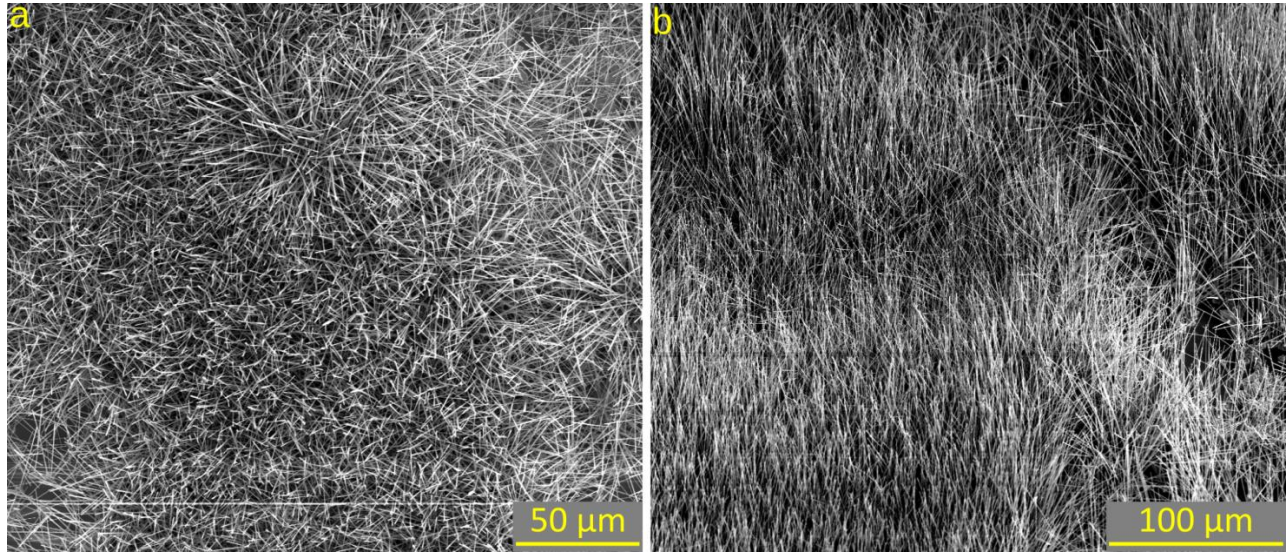

Figure 2. SEM images of ZnO NWs taken at 0° (a) and 45° (b) tilt of the samples.

ZnO NWs were examined using a Titan 80-300 TEM equipped with a spherical aberration image corrector. In order to conduct a TEM examination, NWs were transferred mechanically onto a 3 mm diameter TEM copper grid covered by a holey carbon film. In Figure 3a one can observe an almost atomically smooth edge of the NW. In Figure 3b, we present a tip of the NW. Both Figure 3a and Figure 3b HR-TEM images show the ZnO wurtzite structure with no visible lattice defects, despite the fact that in the first case a thick piece of a NW is shown, and in the second case a thin tip.

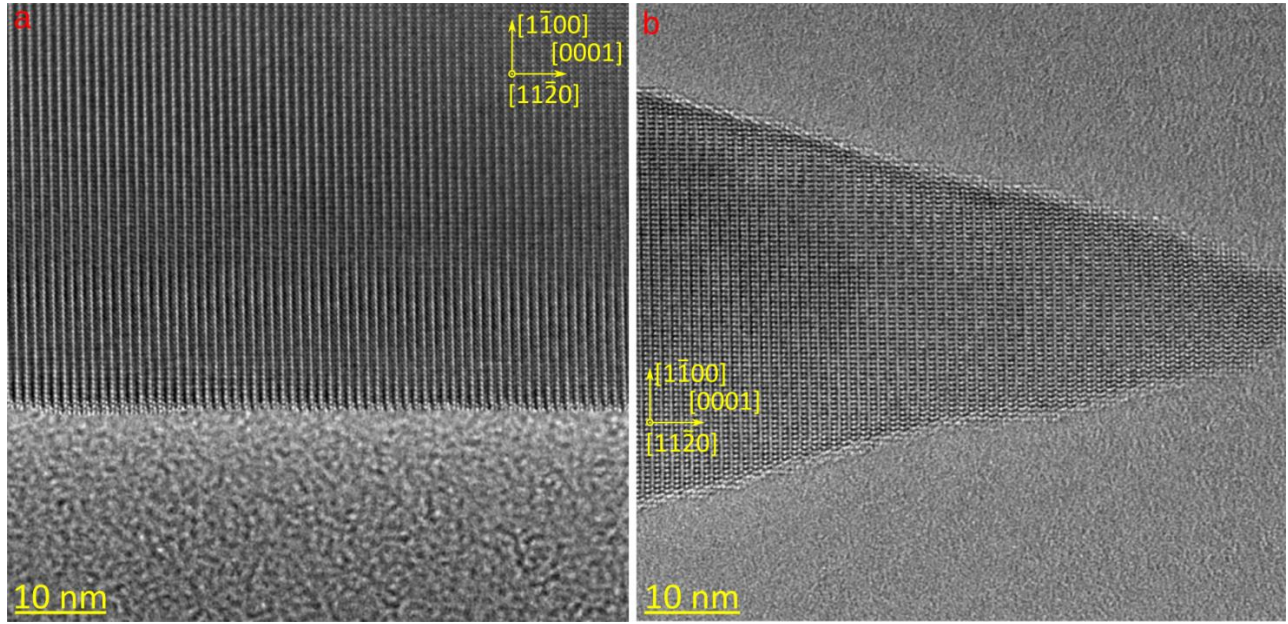

Figure 3 HR TEM images presenting the perfect crystallographic quality of ZnO NWs in both thick (a) and thin (b) cases.

In addition to observing the NWs morphology by SEM, the Focused Ion Beam (FIB) was used to prepare a thin longitudinal section of NWs (simultaneously, cross section of the substrate-spinel ensemble) in order to examine the $ZnAl_2O_4$ layer and two interfaces: $Al_2O_3/ZnAl_2O_4$ and $ZnAl_2O_4/ZnO$. STEM image (Figure 4) was acquired in the Z-contrast with camera length equal to 7,3 cm using a high-angle annular dark-field (HAADF) detector. In Figure 4a a piece of sapphire is visible at the bottom, ZnO at the top and between them there is a buffer layer about 6 nm thick. The layer was identified as $ZnAl_2O_4$ spinel, the orientation of which seems to allow for ZnO NWs vertical growth. The epitaxial relations are as follows: [11-20] $Al_2O_3$||[111] $ZnAl_2O_4$||[0001] ZnO. The reconstruction of the $Al_2O_3$-$ZnAl_2O_4$-ZnO nanostructure was prepared using Crystal Maker (version 10.7.1.300) software and it is presented in two crystallographic orientations, in which zone axis directions are mutually rotated by 90° (Figure 4b,c). Looking along [11-20] direction of ZnO and [2-1-1] direction of spinel, one can observe that the position of every zinc atom from ZnO almost corresponds to the position of the zinc atom from spinel on the interface (Figure 4b, pink frame). Figure 4c presents [1-100] ZnO and [0-11] $ZnAl_2O_4$ view. In this projection, between every two zinc atoms of the spinel there are three zinc atoms of the ZnO NW in the interface (indicated with pink frame, Figure 4c). The lattice mismatch of the spinel in the [111] direction and the ZnO in the [0001] direction was determined to be 2.09%, with the lattice constant of the growing NW being greater.

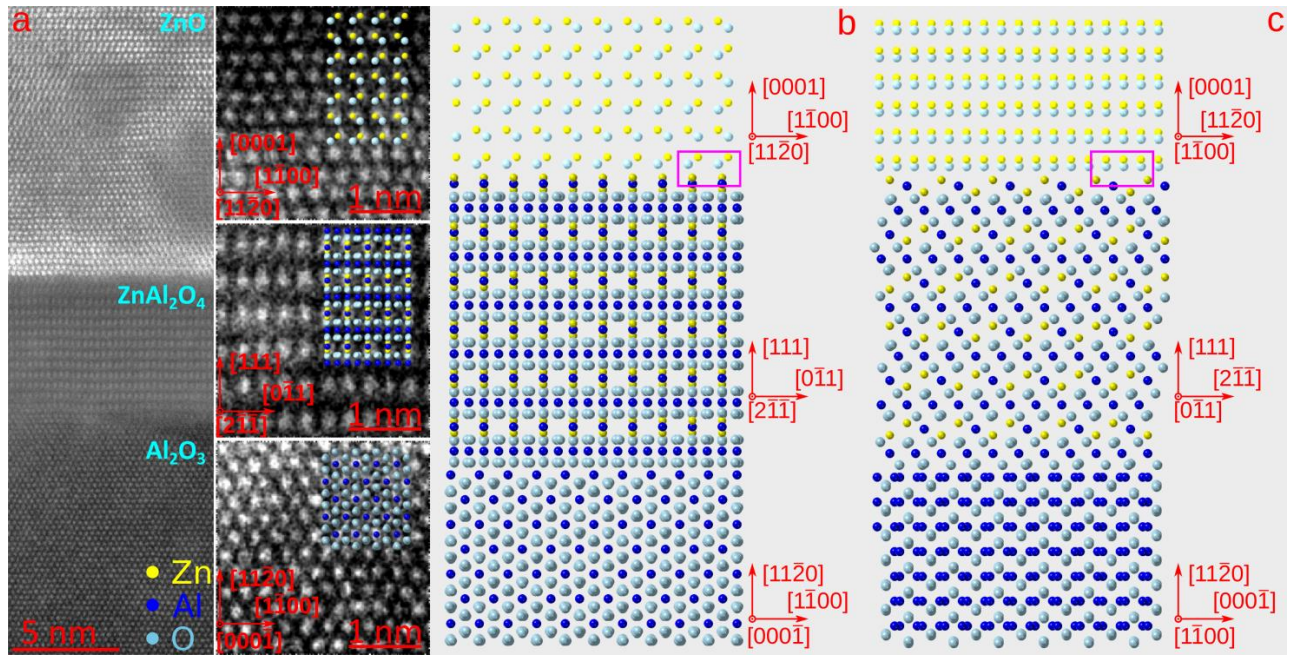

Figure 4. STEM image of $Al_2O_3$-$ZnAl_2O_4$-ZnO NWs structure (a), atomic reconstruction of the structure in two orientations (b,c).

In addition, we present a low-magnification STEM image acquired in the Z-contrast with camera length equal to 7,3 cm using a HAADF detector, similarly to Figure 4a. In [-220-1] $Al_2O_3$ orientation, the spinel layer is visible out of the zone axis. Rotating the specimen by 23.77° vertically and 1.92° horizontally enabled us to observe the $ZnAl_2O_4$-ZnO interface (Figure 5a).

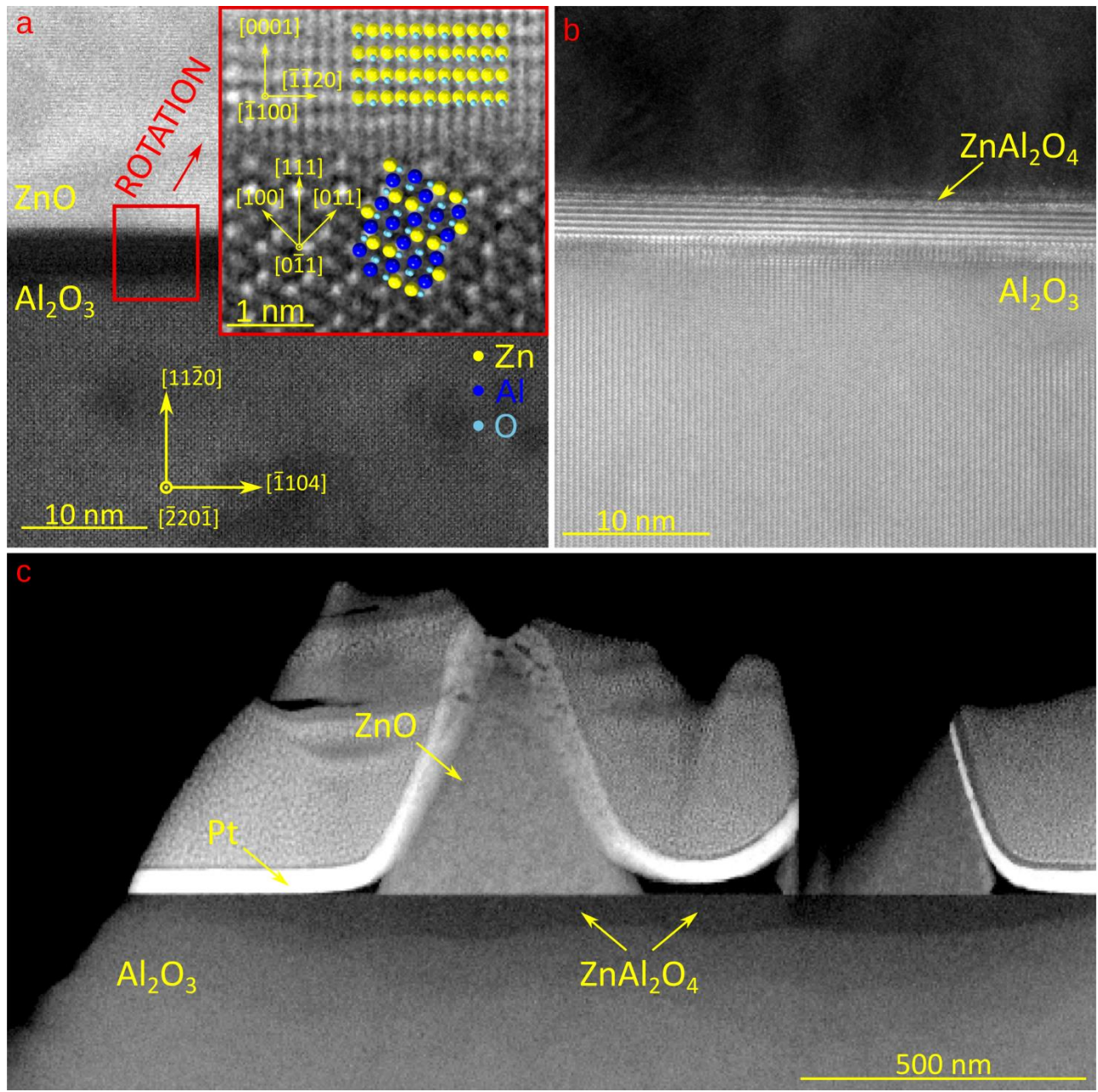

Figure 5. STEM image presenting Al$_2$O$_3$-ZnAl$_2$O$_4$-ZnO NWs structure with zoomed rotated spinel-ZnO NW interface (a), STEM image of Al$_2$O$_3$-ZnAl$_2$O$_4$ structure originating from sapphire-crucible contact zone with no ZnO NWs (b). Low-magnification STEM image of the structure: the spinel is visible throughout the observed sapphire region: with and without ZnO NWs on the top (c).

Two spinel growth mechanisms were considered: ZnO NWs grow on the sapphire substrate firstly, then the spinel layer grows as a product of solid-solid Al$_2$O$_3$- ZnO reaction, or the spinel crystalizes as product of solid-vapor $Al_2O_3$-ZnO reaction firstly, then ZnO NW grows subsequently. To distinguish that, additional lamella was prepared using FIB technique (Figure 5b). The specimen was cut with Ga ions from regions of the sapphire substrate where the ZnO vapors were able to be transferred and take part in solid-vapor $Al_2O_3$–ZnO reaction, but the ZnO NWs could not have grown. This specific area of the substrate is located in the region of the sapphire, where it has contact point with the alumina crucible. This sapphire-crucible contact zone results from fact, that the substrate rests on the shelfs cut in the crucible walls, see Figure 1.

We believe that the specific, threshold conditions must be fulfilled to enable ZnO NWs growth. Sensitivity of the process largely depends on ZnO gas partial pressure, which was observed during ZnO NWs growth. Even small change of the ZnO vapor partial pressure (caused by factors such initial mass of the ZnO:C powder or degree of coverage) influences strongly on density, height and width of obtained ZnO nanostructures. Particularly, too low amount of ZnO vapor results in no ZnO NWs crystallization. Nevertheless, due to specific design of the set-up, there is a region where the amount of ZnO gas is too low to start ZnO NWs process growth (see Figure 1). Diffusion of ZnO vapor along the contact surface is possible because of the spinel appearance.

As mentioned above, there are two possible ways how the spinel theoretically could grow:

1. The ZnO NWs crystallize on $Al_2O_3$ surface firstly, then solid-solid reactions take place, as a result of which a layer of spinel is formed (excluded). If this case were true, we should be able to observe ZnO residues on the spinel on a specimen prepared of the covered region (sapphire-crucible contact zone).
2. The spinel grows on the surface of $Al_2O_3$ as a result of the reaction of the substrate with ZnO vapor, then ZnO NWs grow on the spinel (our case). The sapphire-spinel interface is sharper than spinel-ZnO NWs one – this fact indicates that spinel creation took place in the sapphire: starting from the surface and finished not perfectly regularly (the spinel reached different depths in particular regions of the substrate, see Figure 5c). Then, ZnO NWs started to grow on the smooth spinel surface. In addition, low-magnification image (Figure 4c) presents longitudinal section of sapphire-spinel-ZnO NWs-platinum ensemble (platinum originates from FIB preparation). One can observe, that spinel layer is clearly visible both under ZnO NWs and regions between them. If the first option were true, then we should see a difference in the thickness of the spinel layer under the ZnO NW and the area between the NWs.

We propose the solid-vapor reaction, which leads to $ZnAl_2O_4$ spinel growth:

$$ZnO_{(g)} + Al_2O_{3(s)} \rightarrow ZnAl_2O_{4(s)} \qquad (1)$$

We assume that the reaction (1) takes place on the $Al_2O_3$ surface and the spinel product grows in the depth of the sapphire, as $Al_2O_3$ is a reagent in the process.

4. Conclusions

We have shown a technique which enables us to produce $Al_2O_3$-$ZnAl_2O_4$-ZnO NWs structures. We conclude that the spinel layer grows due to the solid -vapor reaction on the sapphire surface in ZnO vapors and afterwards, ZnO NW grows on the spinel buffer. Examination of a cross sectional low magnification TEM images of $Al_2O_3$-$ZnAl_2O_4$-ZnO NWs ensemble shown that the spinel is present both in case of ZnO NW on top and without it. In addition, $Al_2O_3$-$ZnAl_2O_4$ structure originating from sapphire-crucible contact zone indicates that no solid ZnO was necessary for the spinel growth. In this specific area only partial quantity of gaseous ZnO was diffused between crucible and the substrate resting on it. The presented results prove the good quality of obtained materials and [11-20] $Al_2O_3$||[111] $ZnAl_2O_4$||[0001] ZnO epitaxial relationship was found. Moreover, there is given a model of the found relationship view of the 90° rotated structure. The presented spinel layer has smooth interface with ZnO NW (or creates a smooth surface in areas between NWs) and rough contact surface with Al2O3.

In contrary to our case, in C. Gorla's et.al. examination[16] reactions of the spinel growth concerned solid ZnO - solid $Al_2O_3$. Due to the oxygen sublattice rearrangement at both reaction fronts, there occurred various epitaxial relationships of spinel with $Al_2O_3$ and ZnO. In the H. Fan et. al. article[19], the authors presented experiment of solid-vapor reaction of ZnO NWs with alumina vapor. They did not managed to obtain ZnAl2O4 nanowires, but rippled wires of a ZnO phase.

Appendices

This work has been supported by the National Science Center Poland, through projects No: 2019/35/B/ST5/03434 and 2020/37/B/ST8/03446

References

[1] T. Sahoo, S.K. Tripathy, Y.T. Yu, H.K. Ahn, D.C. Shin, I. H. Lee, Morphology and crystal quality investigation of hydrothermally synthesized ZnO micro-rods. Mater. Res. Bull. **43**, 2060–2068 (2008). DOI: https://doi.org/10.1016/j.materresbull.2007.09.011

[2] M. Willander, Q.X. Zhao, Q.H. Hu, P. Klason, V. Kuzmin, S.M. Al-Hilli, O. Nur, Y.E. Lozovik, Fundamentals and properties of zinc oxide nanostructures: Optical and sensing applications. Superlattices and Microstructures **43**, 352–361 (2008). DOI: https://doi.org/10.1016/j.spmi.2007.12.021

[3] K. Matuła, Ł. Richter, W. Adamkiewicz, B. Åkerström, J. Paczesny, R. Hołyst, Influence of nanomechanical stress induced by ZnO nanoparticles of different shapes on the viability of cells. Soft Matter **12**, 4162–4169 (2016). https://doi.org/10.1039/C6SM00336B

[4] R.Y. Hong, J.H. Li, L.L. Chen, D.Q. Liu, H.Z. Li, Y. Zheng, J. Ding, Synthesis, surface modification and photocatalytic property of ZnO nanoparticles. Powder Technol. **189,** 426–432 (2009). https://doi.org/10.1016/j.powtec.2008.07.004

[5] J. Cui, Zinc oxide nanowires. Mater. Charact. **64,** 43–52 (2012). https://doi.org/10.1016/j.matchar.2011.11.017

[6] L.Z. Kou, W.L. Guo, C. Li, Piezoelectricity of ZNO and its nanostructures, 2008 Symposium on Piezoelectricity, Acoustic Waves and Device Applications, 354-359, China 2008. DOI: https://doi.org/10.1109/SPAWDA.2008.4775808

[7] X. Li, Z. Zhu, Q. Zhao, L. Wang, Photocatalytic degradation of gaseous toluene over ZnAl$_2$O$_4$ prepared by different methods: A comparative study. J. Hazard. Mater. **186,** 2089–2096 (2011). DOI: https://doi.org/10.1016/j.jhazmat.2010.12.111

[8] L. Cornu, M. Gaudon, V. Jubera, ZnAl$_2$O$_4$ as a potential sensor: variation of luminescence with thermal history. J. Mater. Chem. C **1**, 5419-5428. DOI: https://doi.org/10.1039/C3TC30964A

[9] E. Chikoidze, C. Sartel, I. Madaci, H. Mohamed, C. Vilar, B. Ballesteros, F. Belarre, E. del Corro, P. Vales-Castro, G. Sauthier, L. Li, M. Jennings, V. Sallet, Y. Dumnot, A. Perez-Tomas, P-Type Ultrawide-Band-Gap Spinel ZnGa2O4: New Perspectives for Energy Electronics. Cryst. Growth Des. **20**, 2535–2546 (2020), DOI: https://doi.org/10.1021/acs.cgd.9b01669

[10] M. Hoppe, O. Lupan, V. Postica, N. Wolff, V. Duppel, L. Kienle, I. Tiginyanu, R. Adelung, ZnAl$_2$O$_4$-Functionalized Zinc Oxide Microstructures for Highly Selective Hydrogen Gas Sensing Applications. Phys. status solidi **215**, 1700772 (2018). DOI: https://doi.org/10.1002/pssa.201700772.

[11] M. Nasr, R. Viter, C. Eid, F. Warmont, R. Habchi, P. Miele, M. Bechelany, Synthesis of novel ZnO/ZnAl$_2$O$_4$ multi co-centric nanotubes and their long-term stability in photocatalytic application. RSC Adv. **6**, 103692–103699 (2016). DOI: https://doi.org/10.1039/C6RA22623J

[12] B.E. Azar, A. Ramazani, S.T. Fardood, A. Morsali, Green synthesis and characterization of ZnAl2O4@ZnO nanocomposite and its environmental applications in rapid dye degradation. Optik **208**, 164129 (2020). DOI: https://doi.org/10.1016/j.ijleo.2019.164129

[13] O. Lupan, V. Postica, J. Grottrup, A.K. Mishra, N.H. de Leeuw, J.F. Carreira, J. Rodrigues, N. Sedrine, M.R. Correia, T. Monteiro, V. Cretu, I. Tiginyanu, D. Smazna, Y.K. Mishra, R. Adelung, Hybridization of Zinc Oxide Tetrapods for Selective Gas Sensing Applications. ACS Appl. Mater. Interfaces **9**, 4084-4099 (2017). DOI: https://doi.org/10.1021/acsami.6b11337

[14] J.S. Na, Q. Peng, G. Scarel, G.N. Parsons, Role of Gas Doping Sequence in Surface Reactions and Dopant Incorporation during Atomic Layer Deposition of Al-Doped ZnO. Chem. Mater **21**, 5585 (2009). DOI: https://doi.org/10.1021/cm901404p

[15] G. Jáger, J.J. Tomán, L. Juhász, G. Vecsei, Z. Erdélyi, C. Cserháti, Nucleation and growth kinetics of ZnAl$_2$O$_4$ spinel in crystalline ZnO – amorphous Al$_2$O$_3$ bilayers prepared by atomic layer deposition. Scr. Mater. **219**, 114857 (2022). DOI: https://doi.org/10.1016/j.scriptamat.2022.114857

[16] C.R. Gorla, W.E. Mayo, S. Liang, Y. Lu, Structure and interface-controlled growth kinetics of ZnAl$_2$O$_4$ formed at the (11-20) ZnO/(01-12) Al$_2$O$_3$ interface. J. Appl. Phys. **87**, 3736–3743 (2000). DOI: https://doi.org/10.1063/1.372454

[17] J. Zhou, J. Liu, X. Wang, J. Song, R. Tummala, N.S. Xu, Z.L. Wang, Vertically Aligned Zn$_2$SiO$_4$ Nanotube/ZnO Nanowire Heterojunction Arrays. Nano Micro Small **3**, 622-626 (2007). DOI: https://doi.org/10.1002/smll.200600495

[18] Y.J. Li, M.Y. Lu, C.W. Wang, K.M. Li, L.J. Chen, ZnGa$_2$O$_4$ nanotubes with sharp cathodoluminescence peak. Appl. Phys. Lett. **88**, 2–5 (2006). DOI: https://doi.org/10.1063/1.2191418

[19] H.J. Fan, A. Lotnyk, R. Scholz, Y. Yang, D.S. Kim, E. Pippel, S. Senz, D. Hesse, M. Zacharias, Surface reaction of ZnO nanowires with electron-beam generated alumina vapor. J. Phys. Chem. C, **112**, 6770–6774 (2008). DOI: https://doi.org/10.1021/jp712135p

Correspondence address: Wiktoria Zajkowska, Al. Lotników 32/46, 02-668 Warsaw, +14 22 116 3181, zajkowska@ifpan.edu.pl